\documentstyle[aps,prb,epsf,floats]{revtex}

\newcommand{\ve}{\varepsilon}

\begin{document}
\draft

\twocolumn[\hsize\textwidth\columnwidth\hsize\csname@twocolumnfalse%
\endcsname

\title{
Bias and temperature dependence of the 0.7 conductance anomaly 
in Quantum Point Contacts}
\author{
A.~Kristensen, H.~Bruus, A.E.~Hansen, J.B.~Jensen, 
P.E.~Lindelof, C.J.~Marckmann,
J.~Nyg{\aa}rd, and C.B.~S{\o}rensen}   
\address{Niels Bohr Institute, {\O}rsted Laboratory,
  Universitetsparken 5, DK-2100 Copenhagen} 
\author{F.~Beuscher , A.~Forchel, M.~Michel}
\address{ Technische Physik, Universit\"at W\"urzburg, Am Hubland, D-97074 W\"urzburg}

\date{Submitted to Phys.~Rev.~B, May~3, 2000}
\maketitle
\begin{abstract}

The 0.7~$(2e^2/h)$ conductance anomaly is studied in strongly
confined, etched GaAs/GaAlAs quantum point contacts, by measuring the
differential conductance as a function of source-drain and gate bias
as well as a function of temperature. We investigate in detail how,
for a given gate voltage, the differential conductance depends on the
finite bias voltage and find a so-called self-gating effect, which we
correct for. The 0.7 anomaly at zero bias is found to evolve smoothly
into a conductance plateau at 0.85~$(2e^2/h)$ at finite 
bias. Varying the gate voltage the transition between the 1.0  and the
0.85~$(2e^2/h)$ plateaus occurs for  definite bias voltages, which
defines a gate voltage dependent energy difference $\Delta$. This
energy difference is compared with the activation temperature $T_a$
extracted from the experimentally observed activated behavior of
the 0.7 anomaly at low bias. We find $\Delta = k_BT_a$ which lends 
support to the idea that the conductance anomaly is due to
transmission through two conduction channels, of which the one with
its subband edge $\Delta$ below the chemical potential becomes
thermally depopulated as the temperature is increased. 

\end{abstract}

]

\section{Introduction}
\label{sec:introduction}

The quantized conductance through a narrow quantum point contact (QPC),
discovered in 1988\cite{vanWees88,Wharam88}, is one of the key
effects in mesoscopic physics. The quantization of the conductance in
units of the spin degenerate conductance quantum, $G_2 = 2\:e^2/h$,
can be explained within a single-particle Fermi-liquid picture in
terms of the Landauer-B\"{u}ttiker formalism as, in the most simple
case, adiabatic transport through the constriction.
For a review see Ref.~\onlinecite{vanHouten92}. 

Since 1995 several experiments
\cite{Tarucha95,Yacobi96,Tscheuschner96,Thomas96} on quantum wires and
point contacts have revealed deviations from this integer quantization,
$G = n\: G_2, n=1,2,3,\ldots$. In particular the 0.7 conductance
anomaly, noted for the first time in 1991\cite{Patel91} but first
studied in detail in 1996\cite{Thomas96}, poses one of the most
intriguing and challenging puzzles in the field both experimentally
and theoretically 
\cite{Thomas98,Liang99,Kristensen98a,Kristensen98b,Kristensen98c,Reilly00,Berggren,Spivak,Bruus00}.
This anomaly is a narrow plateau, or in some cases just a
shoulder-like feature, clearly visible at the low density side of
the first conductance plateau in the dependence of the conductance $G$
on a gate voltage which tunes the width and the electron density of
the QPC. For low bias voltage the conductance value of the anomalous
plateau is around $0.7\:G_2$ giving rise to the name of the
phenomenon. The 0.7 anomaly has been recorded in many QPC transport
experiments involving different materials, geometries and measurement
techniques.

In this paper, we present experimental evidence, that the 0.7
conductance anomaly is associated with a density-dependent energy
difference separating two transmission channels. We reach this
conclusion by measuring both the temperature and the source-drain bias
voltage dependence of the differential conductance, $G = dI/dV_{sd}$,
through shallow-etched QPCs.

The outline of the paper is as follows. In Sec.~II we describe the
fabrication of the six samples to be investigated. In the following
all detailed results on the conductance of the QPCs are shown solely
for sample A, and only 
towards the end of the paper the main results from all samples are
shown. In Sec.~III we discuss the lateral confinement potential
defining the QPC, and we focus in particular on the fact that this
potential is controlled by two independent variables: the gate bias
and the source-drain bias. Then follows in Sec.~IV the results 
from finite source-drain bias  spectroscopy, and the important energy
difference $\Delta$ is introduced. We deal with the temperature
dependence of the zero-bias conductance in Sec.~V and introduce the
activation energy $T_A$. The main result is obtained in Sec.~VI where
we show that $\Delta = k_B T_A$ for all six samples. A short
conclusion is given in Sec.~VII.

\section{The shallow etched samples}
\label{sec:experimental}

The quantum point contacts were all fabricated on modulation doped
GaAs/GaAlAs heterostructures grown by molecular beam epitaxy (MBE). 
The layer sequence is: 1 $\mu$m ${\rm GaAs}$ buffer, $20 \, {\rm nm}$
${\rm Ga_{0.7} Al_{0.3} As}$ spacer, $40 \, {\rm nm}$ 
${\rm Ga_{0.7}Al_{0.3} As}$ barrier layer with a Si 
concentration of $2 \times 10^{24}$~m$^{-3}$, and a 10~nm undoped GaAs
cap layer. The carrier density is $2\times 10^{15}$~m$^{-2}$ and the
mobility is 100~m$^2$/Vs, measured in the dark at a temperature of
4.2~K.

\begin{figure}[t]
\centerline{\epsfxsize=\columnwidth\epsfbox{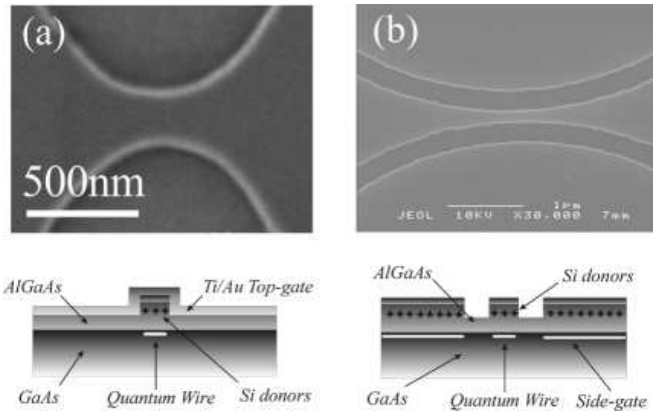}}
\narrowtext
\caption{\label{fig:sempictures}
Scanning electron microscope pictures of the shallow etched quantum
point contacts. 
(a) Type I devices. The quantum point contact is formed by shallow
wet-etching, $\sim$60~nm deep. The etched walls are shaped as two
back-to-back parabolas. The picture was recorded before covering the
etched constriction with a 10$\mu$m wide, 100nm thick Ti/Au top-gate. 
(b) Type II and III devices. Two semicircular shaped, etched trenches
define the quantum point contact and two large areas of 2DEG, which
are used as side-gates. 
In type II devices, the trenches are etched 60nm deep to remove the
donor layer. 
In type III devices the trenches are etched 90nm to the GaAlAs/GaAs
heterointerface, and subsequently covered with GaAlAs by MBE
re-growth. 
}
\end{figure}

The samples were processed with a $20 \times 100$~($\mu$m)$^2$ mesa,
etched 100~nm, and AuGeNi ohmic contacts to the 2DEG were formed
by conventional UV-litho\-gra\-phy, lift-off and annealing. The narrow
QPC constriction was defined using electron beam lithography 
(EBL) and shallow wet-etching on the mesa. 
The following procedure was used: The sample was flushed in acetone,
methanol and iso-propanol before it was ashed in an oxygen plasma for
20 seconds. 
The sample was then pre-etched in 18\% HCl for 5 minutes,
flushed in H$_2$O and blown dry in nitrogen. It was then
pre-baked for 5 minutes at 185~$^{\circ}$C before spinning 
on a 125~nm thick layer of PMMA electron beam resist. 
The EBL pattern was exposed with an acceleration voltage of 30~kV, and
developed in MIBK:iso-propanol (1:3). 
The sample was post-baked for 5 minutes at 115~$^{\circ}$C, and ashed
6~seconds before etching 55-60~nm in ${\rm H_{2}O : H_{2}O_{2} :
H_{3}PO_{4}}$ $(38:1:1)$ at an etch rate of 100~nm/min.

Three types of devices were investigated: top-gated (type~I), 
side-gated (type~II), and overgrown side-gated
(type~III). Fig.~\ref{fig:sempictures}a shows a scanning electron 
microscope (SEM) picture of a type~I QPC constriction. 
The shallow 
etched walls of the constriction are shaped as two back-to back
parabolas. The picture was taken before the constriction was covered
by a 10~$\mu$m wide, 100~nm thick Ti/Au top-gate 
electrode. In type~II devices, Fig.~\ref{fig:sempictures}b, the QPC
constriction is formed by etching two semi-circular trenches,
$\sim$250~nm wide and $\sim$60~nm deep. The etched trenches also
define two large areas of 2DEG which are used as side-gates. The same
pattern is used in type~III devices, but the trenches are etched 90~nm
deep to reach the GaAs/GaAlAs interface and then MBE-regrown. In this
way the constriction is bounded by heterostructure-interfaces, both
vertically and laterally. The e-beam patterning and the MBE-regrowth
was made before the Ohmic contacts were deposited. Before the
regrowth, the sample was desorbed at $630\:^{\circ}$C for 2~minutes in
the MBE-chamber. The sample was then overgrown with 100~nm undoped
${\rm Ga_{0.9}Al_{0.1}As}$ and a 5~nm undoped GaAs cap layer, using a 
growth temperature of $590\:^{\circ}$C. The sample parameters are
tabulated in Table~\ref{table:samples}. 

The samples were mounted in a liquid helium refrigerator, and the
differential conductance, $G = dI_{\rm sd}/dV_{\rm sd}$, was measured
with a small ac excitation voltage, 5-50~$\mu$V rms,  using
standard lock-in techniques at 33-117 Hz. The effective width of the
QPC and the electron density inside it is controlled by a gate
voltage, which is applied between the source contact and the top or
side gate electrode. Henceforth this gate voltage is denoted 
$V_{\rm gs}$.

\begin{table}[t]
\caption{\label{table:samples}
The six quantum point contact samples investigated in this
paper. Three types of devices were processed from the same modulation
doped GaAs/GaAlAs heterostructure: (I) shallow etch and top-gates,
(II) shallow etched trenches, and (III) shallow etched trenches and
MBE regrowth. Geometry related parameters are shown together with the
first subband spacing $\Delta_{01}$.}  
\begin{center}
\begin{tabular}{l|c|c|c|c|c|c}
Sample     &   A    &    B     &   C  &    D  &   E   &     F  \\ \hline
type       &  I  & II & II & II & III & III \\ \hline
width (nm)  &  200& 150 & 140  & 110 & 100 & 100 \\ \hline
radius ($\mu$m) & 0.1& 2 & 5  & 10 & 2 & 2  \\ \hline
$\Delta_{01}$ (meV) & 6.5 & 7.5 & 9.7 & 10.0 & 5.7 & 5.9 \\
\end{tabular}
\end{center}
\end{table}

\section{The lateral confinement}
\label{sec:confinement}

The shallow etching technique gives rise to a strong lateral
confinement in the constriction. We have previously reported
observation of quantized conductance at temperatures above 30~K in a
50~nm wide shallow etched QPC with a 1D-subband energy separation 
$\Delta_{01} \approx 20$~meV \cite{Kristensen98a}. In this paper our
main example is sample~A (type I), but all the measurements reported
for this sample have also been performed for the others.
Fig.~\ref{fig:gatecharacteristic} shows the gate-characteristics,
{\it i.e.\/} the differential conductance, $G$, as function of 
gate-source voltage, $V_{\rm gs}$, of sample~A, measured at
different temperatures. The 200~nm wide, etched QPC constriction is
depleted at zero gate-voltage, and a positive gate-source voltage is
necessary to open it. We estimate the 1D-subband energy separations in
the QPC's from the thermal smearing of the conductance plateaus, and
more precisely by finite bias spectroscopy as described below.
For the 200~nm wide QPC constriction in device~A we
find an energy separation between the two lowest 1D-subbands,
$\Delta_{01} = 6.5$~meV, see also Table~\ref{table:samples} and
Sec.~\ref{subsec:transcond}. 

The confinement potential $U$ determines the transmission properties
of the device. It is mainly defined by the sample parameters, the
geometry, and the gate-source voltage $V_{gs}$. However, to some extend,
especially near pinch-off where the electron density is low, it does 
also depend on the bias voltage $V_{sd}$ \cite{MartinMoreno}. 
In short we write $U = U(V_{gs},V_{sd})$. This effect of $V_{sd}$
influencing $U$ we denote 'self-gating' since it resembles the
ordinary gate effect from $V_{gs}$ \cite{CLM}. A sample exhibiting a
self-gating can be said to be 'soft', if not it is 'rigid'.

The current $I$ through the QPC can be expressed in terms of
the transmission functions ${\cal T}_n(\ve)$ and the difference
$\Delta f(\ve)$ in thermal occupation factors for the source and drain
reservoirs as:
\begin{figure}[t]
\centerline{\epsfysize=50mm\epsfbox{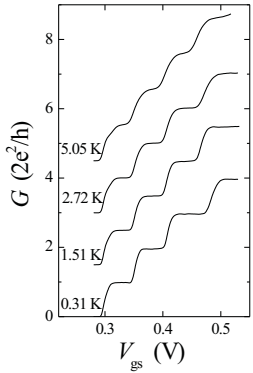}}
\narrowtext
\caption{\label{fig:gatecharacteristic}
Conductance versus gate-voltage at different temperatures measured on
device~A. The strong lateral confinement gives a 1D  subband energy
separation $\Delta_{01} = 6.5$~meV. Well-behaved quantized
conductance plateaus are observed in the temperature range from 300~mK
to above 4~K. 
}
\end{figure}
\begin{equation} \label{eq:LBcurrent}
I = \frac{2e}{h} \sum_n \int_{-\infty}^\infty \! d\ve \: 
{\cal T}_n(\ve) \Delta f(\ve),
\end{equation}
where
\begin{eqnarray}
{\cal T}_n(\ve) &=& {\cal T}_n[\ve,U(V_{gs},V_{sd})]\\
\label{eq:Deltaf}
\Delta f(\ve) &=& f[\ve-\mu-\nu eV_{sd}] -
f[\ve-\mu+(1-\nu) eV_{sd}],
\end{eqnarray}
with $\nu$ being a number between 0 and 1 describing the ratio of the
potential drop on each side of the constriction. Our experimental
results are compatible with $\nu = 1/2$. Writing explicitly the most
relevant functional dependencies for the current we obtain:
\begin{equation} \label{eq:current}
I = I[U(V_{gs},V_{sd}),\Delta f(V_{sd})].
\end{equation}
From this follows to first order in a Taylor expansion the expressions
for the differential conductance $dI/dV_{sd}$ and the transconductance
$dI/dV_{gs}$, the quantities measured in the experiments:
\begin{eqnarray}
\label{eq:dIdVsd}
\frac{dI}{dV_{sd}} &\approx& 
\frac{\partial I}{\partial U} \frac{\partial U}{\partial V_{sd}} +
\frac{\partial I}{\partial \Delta f} 
\frac{\partial \Delta f}{\partial V_{sd}}, \\
\label{eq:dIdVgs}
\frac{dI}{dV_{gs}} &\approx& 
\frac{\partial I}{\partial U} \frac{\partial U}{\partial V_{gs}}.
\end{eqnarray}
We note that any sharp features in the transconductance reminiscent of
the characteristic step-like form in the conductance 
(see Fig.~\ref{fig:gatecharacteristic}) derives from the factor
$\partial I/\partial U$ in Eq.~(\ref{eq:dIdVgs}) relating to the
opening of new conductance channels. The other factor $\partial
U/\partial V_{gs}$ is just varying smoothly due to its origin in
electrostatics over length scales of the order of at least 100~nm. But 
$\partial I/\partial U$ also appears as a prefactor in the first term
of the differential conductance in Eq.~(\ref{eq:dIdVsd}). Thus the
self-gating effect is enhanced when the transconductance is large.
Conversely, at low temperatures at the middle of a
plateau the current is almost unaffected by changes in $U$, at least
only very smooth changes are expected. If $\partial I/\partial U$ can
be neglected, the differential conductance is given by the occupation
factor related second term in Eq.~(\ref{eq:dIdVsd}). As the
temperature is enhanced the transconductance becomes more important
even at the center of the plateau as is evident for the highest
temperatures in Fig.~\ref{fig:07structure}.

\section{Bias spectroscopy and the energy difference $\Delta$}
\label{sec:biasdep}

An important source of information about the energy subbands in a QPC
is finite bias spectroscopy. We use the technique developed by Patel
{\em et al.\/} \cite{Patel91} and described theoretically by Glazman
and Khaetskii\cite{Glazman89}. The differential conductance, 
$G=dI/dV_{\rm sd}$, at finite dc source-drain bias voltage, 
$V_{\rm sd}$ is measured by lock-in technique, using a small ac
signal, 50~$\mu$V~rms 117~Hz, superposed on the dc source-drain bias
voltage.

\subsection{The differential conductance at finite bias}
\label{subsec:diffcond}

\begin{figure}[t]
\centerline{\epsfysize=50mm\epsfbox{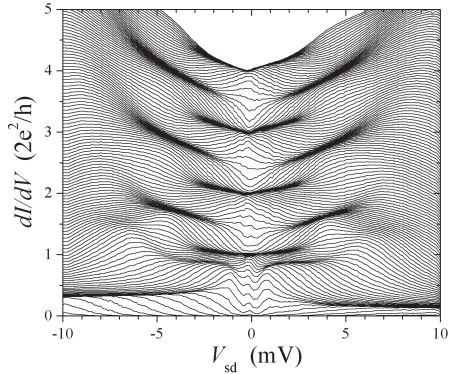}}
\narrowtext
\caption{\label{fig:waterfall}
A plot of the raw data recorded at $T = 0.3$~K of the differential
conductance, $dI/dV$, versus the source-drain bias voltage 
$V_{\rm sd}$ for sample A. Each trace shows $dI/dV$ as $V_{\rm sd}$ is
swept from $-10$ to $10$~mV at fixed gate voltage. The gate voltage is
varied in steps of 1~mV. The first four integer conductance plateaus
are clearly seen around the vertical line $V_{sd}=0$. Also the
corresponding half-plateaus are seen for 
$\sim2~{\rm mV}<|V_{sd}| <\: \sim6~{\rm mV}$. A well developed
0.9 plateau is seen for $\sim1~{\rm mV}<|V_{sd}|<\: \sim4~{\rm mV}$
evolving from a rather weak 0.7 anomaly at $V_{sd} \approx 0$~mV. 
Finally, an additional plateau feature is observed at $G \approx 1.4
\: G_2$ for  $\sim6~{\rm mV}<|V_{sd}|<\: \sim8~{\rm mV}$. 
}
\end{figure}

In Fig.~\ref{fig:waterfall} it is shown how at $T=0.3$~K
the differential conductance of sample~$A$ depends on the dc
source-drain bias. For each trace the gate voltage is fixed, while
going from one trace to the next represents an increase in gate
voltage of 1~mV. Conductance plateaus appear as dark regions with a high
density of traces. Four types of plateaus are observed in the
data. (1) The first four integer conductance plateaus are
clearly seen at $G = n \: G_2$ around $V_{\rm sd}=0$. (2) The
corresponding half-plateaus~\cite{Patel90,Kouwenhoven89} 
at approximately $(n-1/2) \: G_2$ appears for bias voltages
$\sim2~{\rm mV}<|V_{sd}|<\: \sim6~{\rm mV}$, when the chemical
potential of one reservoir lies above the edge of one subband, 
while the other potential lies below. (3) We remark that the
0.7 structure is observed observed near $V_{\rm sd} = 0$. As the 
source-drain bias is increased, the $G$-value of the conductance
anomaly increases, and for $|V_{\rm sd}| \approx 1$~mV, the anomaly
has evolved into a well-defined plateau with a conductance $G$ between
0.8 and 0.9$\: G_2$. (4) Finally, an additional plateau feature is
observed at $G \approx 1.4 \: G_2$ for  $\sim6~{\rm mV}<|V_{sd}|<\:
\sim8~{\rm mV}$. 

From the data in Fig.~\ref{fig:waterfall} it is seen that the
differential conductance depends rather strongly on $V_{sd}$. For the
lowest conductances a pronounced asymmetry is observed: for negative
$V_{sd}$ the conductance is higher than for positive $V_{sd}$. This
effect is always seen when the gate bias is applied relative to the
source contact. It persist in all samples even for different grounding
points. Furthermore, even at the smallest source-drain bias we observe
a strong non-linearity in the conductance at the middle of the integer
plateaus, where the chemical potentials lie in the middle of the gap
between 1d subband edges: the integer plateaus in
Fig.~\ref{fig:waterfall} are not flat around $V_{sd}=0$. In the
following we interpret this non-linearity and the asymmetry in terms
of the self-gating effect presented in Sec.~\ref{sec:confinement}. We
subtract this trivial effect from the data to obtain data
corresponding to a 'rigid' QPC not subject to self-gating.  

\begin{figure}[t]
\centerline{\epsfxsize=0.5\textwidth\epsfbox{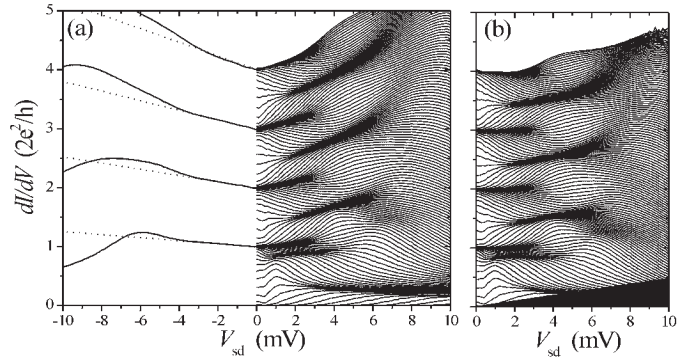}}
\narrowtext
\caption{\label{fig:flat_waterfall}
(a) The symmetrized plot of the differential conductance. In the right
half is shown all the conductance traces, while in the left part is
only shown the four center plateau traces (full lines) together with
the best fit (dotted lines) to the form
Eq.~(\protect\ref{eq:dIdVsd2}). (b) The symmetrized plot 
after subtraction of the $V_{sd}$ dependence due to self-gating.
}
\end{figure}

First we treat the asymmetry of the data, which is most strong for
the lowest values of $V_{gs}$ or equivalently for the lowest electron
densities. A simple reason for this can be found in the electrostatics
of the QPC. We notice that $\frac{\partial I}{\partial U}$ is always
antisymmetric with respect to $V_{sd}$. However, since the gate
voltage is applied relative to the source contact, no special symmetry
relations are expected in $\frac{\partial U}{\partial V_{sd}}$ as the 
polarity $V_{sd}$ is changed. Especially near
pinch-off when the electron density is low in the QPC the effect of a
polarity change in $V_{sd}$ can be important. Thus we expect on
general grounds that regarded as a function of $V_{sd}$ the term
$\frac{\partial I}{\partial U} \frac{\partial U}{\partial V_{sd}}$
from Eq.~(\ref{eq:dIdVsd}) contains both a symmetric and an
antisymmetric part. This conclusion holds true for any value of the
ratio $\nu$ of the voltage drop in Eq.~\ref{eq:Deltaf} in contrast to
Ref.~\onlinecite{MartinMoreno}, where $\nu \neq 1/2$ had to be adopted
to explain the asymmetry. The antisymmetric part thus attributed to
rather trivial electrostatics is subtracted from the data by forming
the symmetric combination
\begin{equation} \label{eq:Isym}
I(|V_{sd}|) \equiv \frac{1}{2}[I(+V_{sd}) + I(-V_{sd})].
\end{equation}

Next we focus on the four $dI/dV_{sd}$ traces which for $V_{sd}=0$
goes right through the center of each of the first four integer
conductance plateaus. As mentioned in Sec.~\ref{sec:confinement}
no appreciable self-gating effect is expected here. Only smooth
changes with $V_{gs}$ is expected for moderate values of the bias
$V_{sd}$. Using a second order Taylor expansion of $dI/dV_{sd}$ in
$V_{sd}$ we extend Eq.~(\ref{eq:dIdVsd}) to the form
\begin{equation} \label{eq:dIdVsd2}
\frac{dI}{dV_{sd}} \approx (\alpha V_{gs} + \beta) +
(\alpha' V_{gs} + \beta') V_{sd},
\end{equation}
and fit the four parameters $\alpha$, $\beta$, $\alpha'$, and $\beta'$
to the four mid-plateau traces. We then subtract from all the traces
the fitted $V_{sd}$ dependence. The result of this procedure is shown
in Fig.~\ref{fig:flat_waterfall}. We end up with plots of the integer
plateaus in the differential conductance which for moderate 
values of $V_{sd}$ up to 2-3~mV are independent of the finite bias
voltage. Note how also the 0.9 anomalous plateau has now become flat. 
We can thus unambiguously assign constant values for the conductance
plateaus in a wide range. The half-plateaus, however,  still show a
dependence of the bias voltage, although not as strongly as before,
indicating the large influence of $V_{sd}$ on the potential $U$ in the
strong non-equilibrium case where one reservoir is injecting electrons
above the topmost subband edges and the other not. We note that
experimentally we never see $G=0.5$ at the first half-plateau but
rather a value substantially below and never quite constant but
decreasing with increasing bias; in the present case $G \approx 0.3$. 
This is probably due to the intricate self-consistent electrostatic
effects at pinch-off, but this have to be investigated further. The
measured values of the conductance at the plateaus are discussed
further in Sec.~\ref{subsec:Delta}.

\subsection{The transconductance}
\label{subsec:transcond}

\begin{figure}[t]
\centerline{\epsfysize=50mm\epsfbox{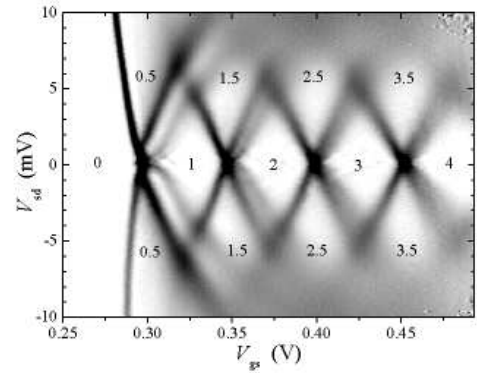}}
\narrowtext
\caption{\label{fig:gate-bias}
Grayscale plot of the transconductance, $dG/dV_{\rm gs}$
versus gate voltage, $V_{\rm gs}$, and bias voltage, $V_{\rm sd}$, for
sample A at $T = 0.3$~K. White corresponds to zero transconductance,
{\em i.e.\/} to plateaus in the differential conductance,
$G=dI/dV_{sd}$. Black corresponds to high transconductance. The dark
lines in the plot therefore indicate the positions $(V_{\rm gs}, 
V_{\rm sd})$ of transitions between the various conductance
plateaus. The numbers indicate the value of $G$ in units of $2e^{2}/h$
on the various plateaus.  
}
\end{figure}

To display the features in the conductance traces more clearly we study
the transconductance, $dG/dV_{\rm gs}$, which is calculated by
numerical differentiation from  the measured differential conductance
$G= dI/dV_{sd}$. The transconductance is zero 
(or small) on conductance plateaus and shows peaks in the transition
regions between plateaus. In Fig.~\ref{fig:gate-bias} is
shown a grayscale plot of the transconductance of sample~A, calculated
from the data in Fig.~\ref{fig:waterfall}. The plot covers the range
$-10$ to 10~mV in source-drain bias and 0.25 to 0.50~V in gate
voltage corresponding to the first four integer conductance plateaus.
Plateau regions (small transconductance) appear as light regions
bounded by dark transition regions (high transconductance). The main
feature of the plot is the well-known diamond shaped dark transition
regions surrounding the integer plateaus $n\:G_2$  and the
half-plateaus  $(n-1/2)\:G_2$, where $n = 1,2,...$
\cite{Thomas98,Kristensen98c}. The transitions in $G$ are due to
the crossing of the chemical potentials $\mu_s$ and $\mu_d$  of the
source and drain reservoirs through the subband edges defining the
transmitting subbands. The procedure described in
Sec.~\ref{subsec:diffcond} to get rid of the $V_{sd}$ dependence of the
plateau values allows for an unambiguous assignment of conductance
values in each of the diamonds of the transconductance plot. The
subband separation $\Delta_{01}$ is extracted from the main diamond
structure by reading off the value of $V_{sd}$ where the straight
black lines surrounding the 1 diamond intersect indicating the
appearance of the next subband. The intersection is at $(V_{gs},\: V_{sd})
= (0.32~{\rm V},\: 6.5~{\rm mV})$. Thus $\Delta_{01}= 6.5$~mV as listed in
Table~\ref{table:samples}.

\subsection{The anomalous subband edge $\Delta(V_{gs})$}
\label{subsec:Delta}

In addition to the main feature the anomalous conductance plateaus are
seen. The most pronounced is the anomalous $G =0.9$ plateau, which
appears in the left-hand side of the $G=1$ diamond between the
leftmost black straight edge and a curved gray anomalous transition
line. Note how the anomalous transition line is continued smoothly
into the $G=1.5$ diamond. Similar, but much weaker, anomalous
structures are seen running inside the 2 diamond continuing into the
2.5 diamond, and inside the 3 diamond continuing into 3.5 diamond.

Just as the black straight lines in the grayscale plot of
Fig.~\ref{fig:gate-bias} are due to the crossing of  $\mu_s$ and
$\mu_d$ through the subband edges  of the transmitting subbands, it is
tempting to also associate a subband edge crossing with the anomalous
transitions. In particular the strong 

\pagebreak

\begin{figure}[t]
\twocolumn[\hsize\textwidth\columnwidth\hsize\csname@twocolumnfalse%
\endcsname
\hfill\epsfxsize=0.44\textwidth\epsfbox{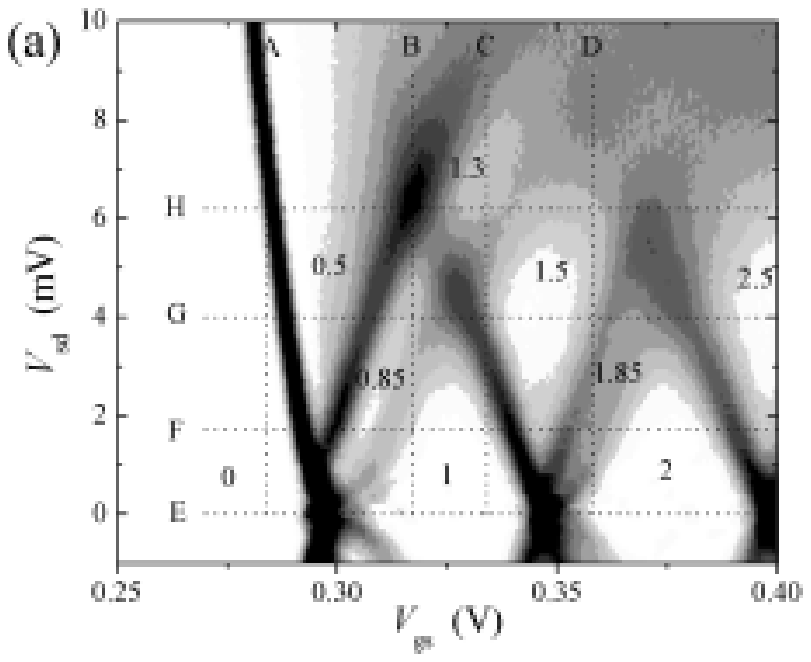}
\hfill\epsfxsize=0.48\textwidth\epsfbox{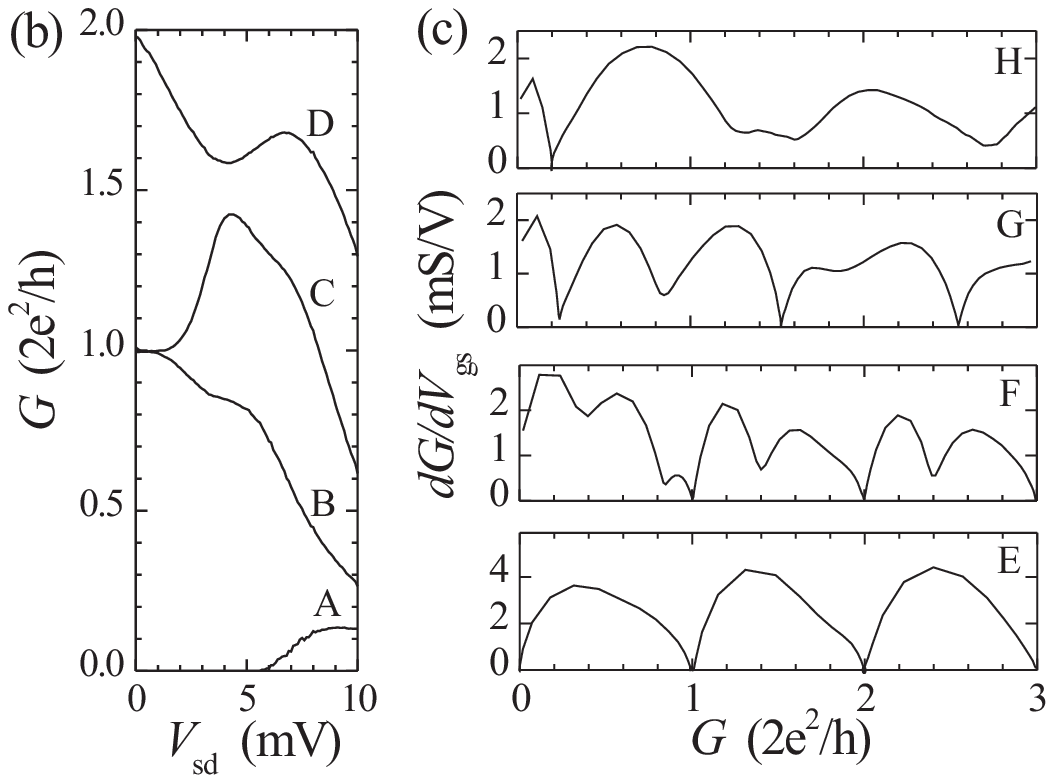}
\widetext
\caption{\label{fig:plateau}
(a) A section of the grayscale plot from
Fig.~\protect\ref{fig:gate-bias} displaying the four vertical
$V_{sd}$ scan-lines A, B, C and D of panel (b) below and the four
horizontal $V_{gs}$ scan-lines E, F, G, and H of panel (c) below. 
(b) The differential conductance $G$ versus bias voltage $V_{sd}$ for
four different fixed values of the gate voltage corresponding to
positions A before the first conductance plateau; B on the lower half
of the same plateau; C on the upper half of it; and D on the lower
half of the second plateau. In panel (c) The transconductance
$dG/dV_{gs}$ versus the differential conductance $G=dI/dV_{sd}$ at
four different bias voltages $V_{sd} = 0$, 1.7, 4.0, and 6.2~mV,
traces E, F, G and H, respectively.
\vspace*{3mm}
}] \narrowtext
\end{figure}

\noindent \mbox{}\\[-10mm]
transition ridge between the 1.0 and the 0.9 plateau can be 
analyzed in those terms. In the standard
theory changing $V_{sd}$ for fixed $V_{gs}$ at the first half of the
first plateau leads to the sequence $G = 1.0 \rightarrow G=0.5$, since
$\mu_d$ drops below the lowest lying spin-degenerate subband 
edge. However, this sequence is not observed in the measurements. 
To make this point clear we show in Fig.~\ref{fig:plateau} four 
individual traces at fixed $V_{gs}$, denoted A to D, and four traces at
fixed $V_{sd}$, denoted E to H. In Fig.~\ref{fig:plateau}a these traces
are drawn as dashed lines in the $V_{gs}$-$V_{sd}$ plane. In
Fig.~\ref{fig:plateau}b is shown the differential conductance along trace
A to D. 
The zero-bias point of these four traces corresponds to the following
positions on the $T=0.3$~K conductance curve of
Fig.~\ref{fig:gatecharacteristic}: below the first plateau (A), on the
lower half of the first plateau (B), on the upper half of the first
plateau (C), and on the lower half of the second plateau (D). First
follow trace B. It exhibits the plateau sequence $G = 1.0
\rightarrow G=0.85 \rightarrow G=0.2$. Probably due to the 'softness'
of the QPC at low electron densities the value of the '0.5-plateau' is
around 0.2, where the trace meet with trace A evolving from $G=0$
into a plateau at $G=0.15$. It is as if the conductance in trace B
drops in two steps corresponding to the crossing of two subband edges
rather than just one, perhaps as a consequence of lifting of the
spin-degeneracy in the QPC\cite{Thomas96,Berggren,Bruus00}.

It seems quite natural to associate the anomalous transition with an 
anomalous subband edge which lies above the ordinary subband edge and
therefore is encountered first as the bias voltage is raised. This
would also account for the continuation of the anomalous transition
into the 1.5 diamond 
as seen by studying the behavior of trace C. Increasing $V_{sd}$ 
from 0 this trace exhibits a clear plateau at 1.0 before it rises and 
develops into a plateau at $G=1.45$ as $\mu_s$ is raised above the
second subband. For slightly larger value of $V_{sd}$ $\mu_d$ falls
below the anomalous subband edge; $G$ drops and the trace
exhibits a shoulder-like feature around $G=1.3$. Only for yet higher
values of $V_{sd}$ does $\mu_d$ drop below the ordinary first subband
leading to $G=1$ and lower values as in the standard case. Thus as a
function of the bias-voltage $V_{sd}$ the plateau sequences $G = 1.0
\rightarrow G=0.5$ and $G = 1.0 \rightarrow G=1.5 \rightarrow G=1.0$,
for the first and second half of the $G=1$-plateau, expected from the
simple half-plateau model, in experiment are seen rather to 
be $G = 1.0 \rightarrow G=0.85 \rightarrow G=0.5$ and
$G = 1.0 \rightarrow G=1.5 \rightarrow G=1.3 \rightarrow G=1.0$.

The values of the conductance at the plateaus are found after the
fitting procedure described in Fig.~\ref{fig:flat_waterfall}. 
The most precise way to obtain these values is through
Fig.~\ref{fig:plateau}c, where the transconductance $dG/dV_{gs}$ is
plotted versus the differential conductance $G$ at four different but
fixed bias voltages, traces E, F, G and H. The plateaus appear as
minima in the curves, since a minimum in the the transconductance
correspond to the point of least slope in plots of $G$ versus
$V_{gs}$. Ideally, if the plateaus are completely flat, the values at
the minima are 0. This happens for example at the integer plateaus seen
in trace E, and the half-plateaus in trace G. The 0.85-plateau is
never completely flat, but in traces F and G it is seen as a well
developed minimum. 

For comparison with the temperature data presented in
Sec.\ref{sec:analysis} we introduce the anomalous gate voltage
dependent (and hence density dependent) energy difference
$\Delta(V_{gs})$. It is related to that particular gate-voltage
dependent value $V_{sd}^*$ of the source-drain bias that maximizes
the transconductance along the 0.9-1.0 and 1.35-1.5 ridges in the
grayscale plot:
\begin{equation} \label{eq:Delta}
\Delta(V_{gs}) = \frac{1}{2}eV_{sd}^*(V_{gs}),
\end{equation}
In terms of an anomalous subband, $\Delta$ is interpreted as the
difference between the chemical potential and the anomalous subband
edge. In Fig.~\ref{fig:gate-bias} it is seen that similar ridges appear,
progressively weaker, for the higher subbands. The weakening of the
effect may be due to less pronounced spin polarization at the higher
densities present when more subbands are occupied
\cite{Berggren}. Finally, we note that in contrast to the normal
plateaus, the anomalous plateaus only appear when both $\mu_s$ and
$\mu_d$ are above a given subband edge: the anomalous plateaus only
appear in the left-hand side of the diamonds in the grayscale
plots. This is another indication that the anomalous plateaus are
related to interaction effects and not simple single-particle subband
effects. 

\begin{figure}[t]
\centerline{\epsfxsize=0.3\textwidth\epsfbox{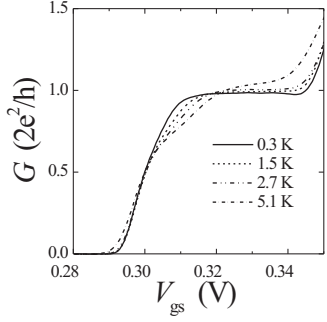}}
\narrowtext
\caption{\label{fig:07structure}
The conductance versus gate voltage of sample~A at the first quantized
conductance plateau, measured at different temperatures from 0.3~K to
5.1~K. As the temperature is raised, the 0.7 anomaly emerges as a
suppression of conductance at the first half of the conductance
plateau, while the other half remains flat. 
}
\end{figure}

\section{The activation temperature $T_a$} 
\label{subsec:tempdep}

To gain further insight in the conductance anomaly we also study the
temperature development of the first conductance plateau, $G = G_2$.
In Fig.~\ref{fig:07structure} is shown a set of measurements
performed on sample~A. At the lowest temperature, 0.3~K, the plateau
is broad and flat. With a 1D subband energy separation of 6.5~meV, the
thermal smearing of the plateau should be negligible at temperatures
below 4~K. This is indeed also the case for the upper half of the
conductance plateau, $V_{gs} \simeq 320-340$~mV, which stays flat as
the temperature is raised. On the lower half of the plateau, the
conductance is suppressed below the plateau value, $G_2$, as the
temperature is raised, developing a plateau-like structure around the
conductance value $0.7 \:G_2$. This is the 0.7 conductance anomaly.

\begin{figure}[t]
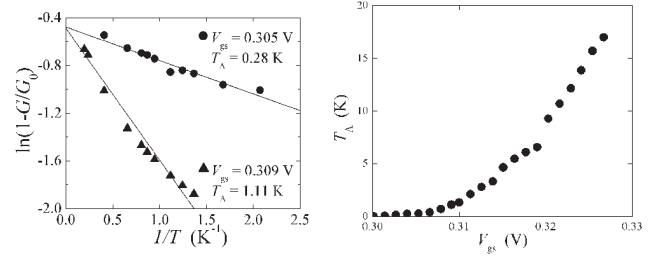

\centerline{\epsfxsize=0.24\textwidth\epsfbox{fig8a.eps}
            \epsfxsize=0.24\textwidth\epsfbox{fig8b.eps}}
\narrowtext
\caption{\label{fig:Arrhenius}
(a) Temperature dependence of the conductance suppression, $G_0-G(T)$,
at fixed gate voltages, $V_{\rm gs}=0.305$~V and 0.309~V, measured on
device~A. The data shows an Arrhenius behavior, $G(T)/G_0=1-C\exp
(-T_{A}/T)$, with an activation temperature, $T_{A}$. (b) The measured
activation temperature, $T_{A}$ as function of gate voltage across the
$0.7\:G_2$ structure, measured on device~A.  
}
\end{figure}

The large 1D-subband energy separation in the shallow etched QPC's
allows us to study the temperature dependence of the $0.7$-structure
at temperatures up to around 5~K without appreciable thermal smearing
of the quantized conductance. In Fig.~\ref{fig:Arrhenius} we present
two Arrhenius plots of the conductance suppression shown in 
Fig.~\ref{fig:07structure} at $V_{gs} = 0.305$~V and 0.309~V. We plot
the relative conductance suppression $1-G(T)/G_0$ (where $G_0$ is the
measured conductance value of the plateau) versus $1/T$ at the given
fixed gate-voltage. The linear behavior in the semilogarithmic Arrhenius 
plot indicate an activated behavior, $G(T)/G_0 = 1 - C\exp(-T_{A}/T)$, 
with the corresponding activation temperatures, $T_{A} = 0.28$~K and
1.11~K, extracted from the two slopes,
respectively. Fig.~\ref{fig:Arrhenius}b shows how the measured
activation temperature $T_A$ as a function of gate voltage increases
from 0 at pinch off to a few kelvin at the middle of the conductance
plateau.

In the usual framework of the Landauer-B\"{u}ttiker formalism the
observed activated suppression of the conductance indicates that the
$0.7$-structure is associated with thermal depopulation of a subband
having a gate voltage dependent subband edge. If a subband edge
lies $k_BT_A(V_{gs})$ below the Fermi level indeed an activated
behavior is seen in $G$. A phenomenological theory along
these lines has been presented by Bruus {\it et al.\/}
\cite{Bruus00}. Moreover, this picture is in accordance with the
discussion presented in Sec.~\ref{subsec:Delta} of the crossing of
subband edges at finite bias. In the following analysis we connect the
measured activation temperature with the energy gap $\Delta$ found by
finite-bias spectroscopy.

\section{Comparing $\Delta$ and $T_a$}
\label{sec:analysis}

\begin{figure}[t]
\centerline{\epsfxsize=0.50\textwidth\epsfbox{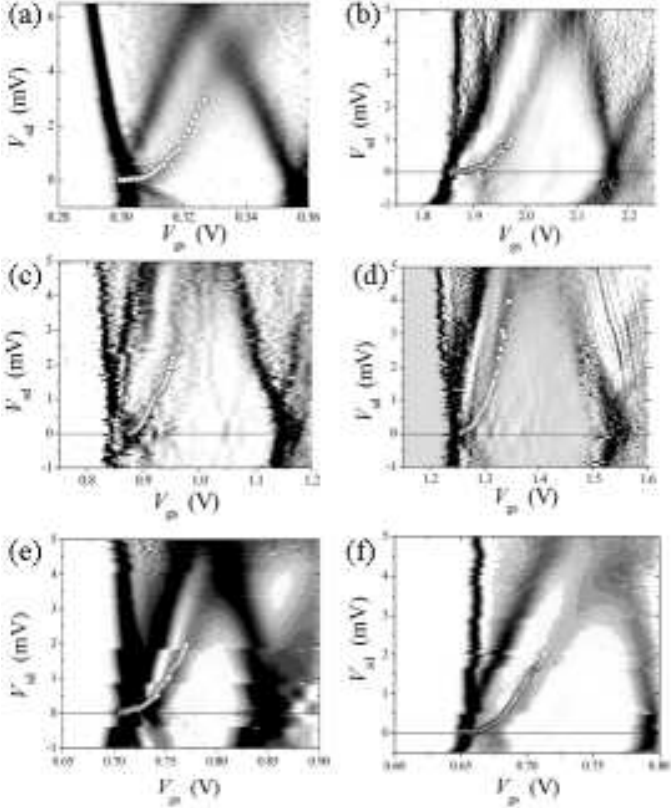}}
\narrowtext
\caption{\label{fig:compare-T-bias} 
Grayscale plots as in Fig.~\protect\ref{fig:gate-bias} of the
differential transconductance for all six samples A-F of
Table~\protect\ref{table:samples}.
The open circles are the data points $(V_{\rm gs} , 2 k_B T_{A}/e)$
where the activation temperature, $T_{A}$ extracted from the 
measured temperature dependence of the zero-bias conductance. 
}
\end{figure}

It is possible to ascribe the same origin to the appearance of the
plateau at $0.9\: G_2$ at finite bias as to the 0.7 anomaly. The two
effects are connected by the energies $\Delta(V_{gs})$ and
$T_A(V_{gs})$. Consider a fixed gate-voltage on the lower half of the
$G_2$ plateau. The data are taken at low temperature.
At zero bias the excitation energies available for the electrons at
the Fermi energy are not sufficient to reach the subband edge lying
$k_BT_A$ below the Fermi level, and the conductance has the expected,
quantized value, $G_2$. As the source-drain bias-voltage, $V_{\rm sd}
= (\mu_{s} - \mu_{d})/e$ is increased we assume that half the
potential drop is before and the other half after the QPC, {\em
  i.e.\/} $\nu = 1/2$ in Eq.~(\ref{eq:Deltaf}). The electrons from the
drain reservoir are injected below the subband edge when $eV_{\rm sd}/2
= k_{B} T_{A}$. This assumption is supported by our experiments. In
Fig.~\ref{fig:compare-T-bias} we have for all six samples plotted the
expected position of the resonance, $V_{\rm sd}^* = 2 k_{B} T_{A}/e$,
versus gate voltage as white circles. The activation temperature used
in this plot is obtained from the measured temperature dependence of the
$0.7$-structure, as the one presented in Figs.~\ref{fig:07structure}
and~\ref{fig:Arrhenius}. As seen from Fig.~\ref{fig:compare-T-bias}
the transition from the regular $G_2$ plateau to the anomalous $0.85\:
G_2$ plateau appears at the expected resonance position.

The quality of the 0.7/0.85 anomalies are varying a lot from sample to
sample. The exact reason for this is not known at present. One can
think of many reasons such as impurities, geometry related defects,
and other sample parameters. But it is noteworthy that for 
{\em all\/} samples the energy $\Delta(V_{gs})$ characterizing the
0.85 anomaly coincides with the activation energy $k_{B} T_{A}$
deduced from the 0.7 anomaly.

\section{Conclusion}
\label{sec:conclusion}
We have investigated the $0.7$ conductance anomaly in six samples of
three different types of shallow etched GaAs/GaAlAS QPC's:  
top-gated, side-gated and side-gated, overgrown. We note that the QPC
confinement potential $U$ depends on both $V_{gs}$ and $V_{sd}$. The
influence from $V_{sd}$, referred to as self-gating, can explain the
distinct asymmetry and non-linearity always observed in differential
conductance of QPCs. The QPCs thus appear to be 'soft', but we have
shown how to subtract the self-gating effect from the data. Based on
finite bias spectroscopy we have presented experimental 
evidence, that the 0.7~anomaly is associated with a density-dependent
energy difference $\Delta$ of the order of a few kelvin being the
distance from the chemical potential to an anomalous subband edge. The
shallow etching technique gives rise to a strong lateral confinement
with 1D subband energy separations of $5-20$~meV. We have therefore
been able study the 0.7 anomaly for higher temperatures than for
normal split-gate devices, and this allowed a detailed study of the
temperature dependence of the conductance anomaly. We have found an
activated behavior of the conductance suppression on the $0.7$
anomaly, with a density-dependent activation temperature, $T_{A}$, of
a few kelvin. For all six samples the energy difference $\Delta$ is
found to be equal to the activation energy $k_BT_A$. Our observations
supports the idea that the 0.7/0.85 conductance anomaly arises from
the existence of an anomalous subband edge in the QPC. The nature of
the anomalous subbands is presently unknown. But our observation that
the anomalous plateaus only appear when both $\mu_s$ and $\mu_d$ are
above a given subband edge, and the behavior of the 0.7 anomaly as
function of magnetic field \cite{Thomas96} indicate the importance of
interaction effects beyond the simple single-particle subband picture,
presumably related to spin polarization \cite{Thomas96,Berggren,Bruus00}.

\section*{Acknowledgements}
\label{sec:acknowledgements}

This research is part of the EU IT-LTR programme Q-SWITCH
(No. 20960/30960), and was 
partly supported by the Danish Technical Research Council (grant
no. 9701490) and by the Danish Natural Science Research Council
(grants no. 9502937, 9600548 and 9601677). 
The III-V materials used in this investigation were made at the
III-V NA\-NO\-LAB, operated jointly by the Microelectronics Centre of the
Danish Technical University and the Niels Bohr Institute
fAPG, University of Copenhagen.

\end{document}